 \documentstyle[12pt]{article}
\begin{document}

 \vspace*{1cm}
 \begin{center}
 {\LARGE \bf Failure of Standard Conservation Laws \\[8mm]
 at a Classical Change of Signature} \\[12mm]

 {\large Charles Hellaby} \\
 {\small
 Department of Applied Mathematics,
 University of Cape Town,
 Rondebosch,
 7700,
 South Africa} \\
 E-mail: {\tt cwh@maths.uct.ac.za} \\[4mm]

 {\large \&} \\[4mm]

 {\large Tevian Dray} \\
 {\small
 Department of Mathematics,
 Oregon State University,
 Corvallis,
 Oregon 97331,
 USA} \\
 E-mail: {\tt tevian@math.orst.edu}

 \vfill

 To appear in {\it Phys. Rev. D} 1994 \\
 gr-qc/9404001 \\

 \vfill

 {\bf \large Abstract} \\[4mm]
 \parbox{12cm}{
     The Divergence Theorem as usually stated cannot be applied across
a change of signature unless it is re-expressed to allow for a finite
source term on the signature change surface.  Consequently all
conservation laws must also be `modified', and therefore insistence on
conservation of matter across such a surface cannot be physically
justified.  The Darmois junction conditions normally ensure
conservation of matter via Israel's identities for the jump in the
energy-momentum density, but not when the signature changes.  Modified
identities are derived for this jump when a signature change occurs,
and the resulting surface effects in the conservation laws are
calculated.  In general, physical vector fields experience a jump in
at least one component, and a source term may therefore appear in the
corresponding conservation law.  Thus current is also not conserved.
These surface effects are a consequence of the change in the character
of physical law.  The only way to recover standard conservation laws
is to impose restrictions that no realistic cosmological model can
satisfy.
 }\\[12mm]


 \vfill

 {\it Figures available on request from Charles Hellaby at above
address.}

 \end{center}

 \vfill

 \newpage

 \noindent {\bf INTRODUCTION}

     Interest in the possibility of a change of the signature of
spacetime has been revived recently, by Hawking's [1] ``No Boundary
Condition" proposal, and by subsequent considerations of quantum
cosmology [e.g. 2,3,4,5,6], and there have been several papers
discussing the junction of Lorentzian (hereafter abbreviated to L) to
Euclidean (hereafter E) regions in classical Relativity
[7,8,9,10,11,12,13].  However, none of these has examined the
Divergence Theorem, upon which all conservation equations are based.
(Stokes' Theorem is discussed in this context in [14] using
differential forms, but conservation is not explicitly discussed.)

     To a large extent, the laws of physics in a space of E
(Euclidean) signature and at a change of signature are a matter of
personal choice - our intuition, which after all is exclusively based
on experience of L (Lorentzian) spacetime, cannot be a reliable guide.
This paper follows a strictly classical approach, which is not
entirely equivalent to the Quantum Cosmology approach, in which the
Euclidean regions are ``classically forbidden".

     We here argue that the Darmois junction conditions (D), which
ensure that the geometries on either side of a boundary surface do in
fact fit together, are the absolute minimum gravitational requirements
for passing through a signature change.   Whilst one may wish to
impose stronger conditions for reasons of preference, or to achieve
some particular physical result, such extra conditions are less
fundamental, and may eliminate legitimate and interesting types of
transition.

     In any case, the primary results obtained here hold for all known
junction conditions, and are independent of the choice of coordinates
near the signature change.  In other words we permit, but do not
assume, a lapse function that goes to zero on the signature change
surface.

     We begin by reviewing the relationship between the
 Darmois-Israel junction conditions [15,16] and the Divergence
Theorem, for the case when no signature change occurs.  We proceed by
considering how the theorem and the Israel identities should be
adjusted for the case when a change of signature does occur inside the
volume of integration.  Similar considerations are applied to an
 electro-magnetic field; and finally the significance of the results
is discussed. \\[1cm]

 \noindent {\bf THE DARMOIS-ISRAEL CONDITIONS}

     We wish to join two manifolds, $M^+$ and $M^-$, of L (Lorentzian)
signature $(-+++)$ with non-null boundary surfaces $\Sigma^\pm$, by
identifying $\Sigma^+$ with $\Sigma^-$.  Manifolds $M^\pm$ have
coordinate systems $x_\pm^\alpha$ and metrics $g^\pm_{\alpha \beta}$,
while $\Sigma^\pm$ have coordinates $\xi^i_\pm$, which are also
identified.  Latin indices range 1 to 3, and Greek indices range 0 to
3.  The Darmois [15] junction conditions (hereafter D) state that the
first and second fundamental forms of the surfaces
 --- the intrinsic metric ${}^3 \!\! g_{ij}$, and the extrinsic
curvature $K_{ij}$
 --- must be continuous across the identified boundary $\Sigma$.
These conditions have been shown to be the ``most convenient and
reliable", whereas those of O'Brien and Synge [17] are too restrictive
in general [18].

     Using the notation
 \begin{equation}  [Z] = \left. Z^+ \right|_\Sigma
       - \left. Z^- \right|_\Sigma   \label{eq:Zjump}  \end{equation}
 for the jump in some quantity $Z$ across $\Sigma$, where $\left.
Z^\pm \right|_\Sigma$ are the limiting values of $Z$ as $\Sigma$ is
approached from either side,
 \begin{equation}  e_i^\alpha = \frac{\partial x^\alpha}{\partial
\xi^i}
   \end{equation}
 for the basis vectors of the surface, and
 \begin{equation}
  n^\alpha \mbox{~~,~~~~~} n^\alpha n_\alpha = \epsilon = \pm 1
\end{equation}
 for the unit normal to $\Sigma$, which may be timelike (i.e. $n$
spacelike, $\epsilon = +1$) or spacelike ($\epsilon = -1$), then the
intrinsic metric and extrinsic curvature are
 \begin{eqnarray}
    {}^3 \!\! g_{ij} & = & g_{\alpha \beta} \: e_i^\alpha \, e_j^\beta
\\
    K_{ij} = \nabla_\beta n_\alpha \: e_i^\alpha \, e_j^\beta  & = &
    - n_\gamma \left( \frac{\partial^2 x^\gamma}{\partial \xi^i
     \partial \xi^j} + \Gamma^\gamma_{\alpha \beta}
      \frac{\partial x^\alpha}{\partial \xi^i}
      \frac{\partial x^\beta}{\partial \xi^j} \right) \mbox{~~.}
              \label{eq:Kij}  \end{eqnarray}
 D are the minimum requirements for joining $M^+$ and $M^-$ smoothly:
 \begin{equation}  [ \, {}^3 \!\! g_{ij}] = 0  \label{eq:Dg}
  \end{equation}
 \begin{equation}  [K_{ij}] = 0  \label{eq:DK}  \end{equation}
 and it is important to note that (\ref{eq:Zjump}) requires the
normals on both sides to point from $M^-$ to $M^+$ for proper
evaluation.  The great advantage of D is that these expressions are
completely invariant to the coordinates used in $M^+$ and $M^-$.  They
may be implemented without ever finding a common coordinate system on
$M^+ \cup M^-$, though one must obviously find a common coordinate
system on $\Sigma^+ = \Sigma^-$.  Thus they provide an unambiguous
algorithm for joining spacetimes.  In the above we have assumed an
isometry between the points on the surfaces $\Sigma^+$ and $\Sigma^-$
induced in $M^+$ and $M^-$.  In simple cases this may merely be an
identification of induced surface coordinates, $\xi^i_+ = \xi^i_-$,
but in general one might have to solve the 3-d metric equivalence
problem before stating whether (\ref{eq:Dg}) may be satisfied or not.
Any isometry for which (\ref{eq:Dg}) and (\ref{eq:DK}) are satisfied
results in a valid matching.

     It is often convenient to use geodesic normal coordinates
(Gaussian coordinates), defined near $\Sigma$ to consist of the proper
time/distance coordinate $\xi^o = \tau$ along geodesics normal to the
surface, increasing from $M^-$ through $\Sigma$ into $M^+$, and the
surface coordinates $\xi^i$ which are held constant along each
geodesic.  Then
 \begin{eqnarray}
    ds^2 & = & \epsilon d\tau^2 + {}^3 \!\! g_{ij} d\xi^id\xi^j
                = \tilde{g}_{\mu \nu} d\xi^\mu d\xi^\nu
       \mbox{~~,~~~~~} \tilde{g}_{ij}|_\Sigma = {}^3 \!\! g_{ij}
\label{eq:ncg}  \\
       \tilde{K}_{ij} & = & - \tilde{\Gamma}_{o,ij} =
\tilde{\Gamma}_{j,oi}
        = \frac{1}{2} \tilde{g}_{ij,o}  \mbox{~~,~~~~~}
\tilde{K}_{ij}|_\Sigma = K_{ij}
                              \label{eq:ncK}  \end{eqnarray}
 where we use a tilde to indicate 4-dimensional quantities expressed
in this Gaussian coordinate system.  Nevertheless, all of the
following may be done without introducing these coordinates.

     Israel [16] has shown that these junction conditions lead to the
following identities for the Einstein tensor, $G_{\alpha \beta}$,
 \begin{equation}  [\tilde{G}_{oo}] = [ G_{\alpha \beta} \: n^\alpha
n^\beta ]
        = 0  \label{eq:Is1}  \end{equation}
 \begin{equation}  [\tilde{G}_{oi}] = [ G_{\alpha \beta} \: n^\alpha
e_i^\beta ]
       = 0  \label{eq:Is2}  \end{equation}
 This means, for a timelike surface, that the flux of energy-momentum
through $\Sigma$, as measured by an observer moving with the surface,
is continuous across $\Sigma$.  For a spacelike surface, an observer
moving orthogonally to it sees no jump in the density of energy-
momentum across $\Sigma$.  However, if only the first fundamental form
is continuous, and there is a jump in the second form, then Israel
showed that $\Sigma$ contains a finite amount of matter and a
``surface layer" occurs.

     D are more or less equivalent to making the appropriate
components of the gravitational field and its first derivatives
continuous across $\Sigma$, naturally expressed in geometric fashion.
Though there are 12 conditions on the 50 independent components of
$g_{\alpha \beta}$ and $g_{\alpha \beta , \gamma}$, there do exist
coordinates in which the 4-d metric and its first derivatives are
continuous [e.g. 19], but these are not always trivial to find
 --- the most reliable choice being normal coordinates.  (This is the
approach of Lichnerowicz [20], which is equivalent to D [18], but the
fact that the Lichnerowicz junction conditions are not invariant makes
them less reliable.)  Nevertheless, even in normal coordinates, the
continuity of all components of the matter tensor $T^{\alpha \beta}$
does not follow from D.  Despite this, conservation of matter right
through the boundary is guaranteed. \\[8mm]

 \noindent {\bf MATTER CONSERVATION AT A BOUNDARY}

     Given a volume $W$ enclosed by a surface $S$ with normal
$m_\alpha$, and defining a 3-form $\bf p$ to have components
 \begin{equation}
  {\bf p_{\alpha \beta \gamma}} = \eta_{\alpha \beta \gamma \delta}
\Psi^\delta =
  \sqrt{\epsilon g} \; \varepsilon_{\alpha \beta \gamma \delta} \;
\Psi^\delta \label{eq:p}
 \end{equation}
 where $\eta_{\alpha \beta \gamma \delta}$ is the permutation tensor
and $\varepsilon_{\alpha \beta \gamma \delta}$ the permutation symbol,
then Stokes' Theorem [e.g. 21] in terms of differential forms
 \begin{equation}
  \oint_S {\bf p} = \int_W d {\bf p}  \label{eq:Stokes}
  \end{equation}
 applies over any region $W$ bounded by $S$ within which $\bf p$ is
$C^1$, and, given a metric, it leads to the Divergence Theorem [e.g.
21] in terms of tensor components
 \begin{equation}  \oint_S \Psi^\beta m_\beta \; d^3S
           = \int_W \nabla_\beta \Psi^\beta \; d^4W   \label{eq:Div}
  \end{equation}
 where $d^4W$ is the metric volume element on W, $d^3S$ is the induced
metric volume element on $S$, and $m^\alpha$, the unit vector normal
to $S$, has contravariant components that point outwards where it
($m^\alpha$) is spacelike, and inwards where it is timelike - i.e.
$m_\alpha$ is always outwards.  (The character of this normal will be
clarified later.)  In general both $\Psi^\alpha$ and $g_{\alpha
\beta}$ must be $C^1$ to make $\bf p$ $C^1$.  In order to give Stokes'
Theorem physical meaning, $\bf p$ must be related to measurable
quantities, which requires a metric.  Hence (\ref{eq:Div}) is the
physical version of (\ref{eq:Stokes}) for a 3-form.  Choosing
 \begin{equation}
  \Psi^\delta = G^{\sigma \delta} v_\sigma   \label{eq:pG}
\end{equation}
 $v_\alpha$ being some smooth field (e.g. an element of an orthonormal
basis), this becomes
 \begin{equation}  \oint_S G^{\alpha \beta} v_\alpha m_\beta \; d^3S
           = \int_W \nabla_\beta (G^{\alpha \beta} v_\alpha) \; d^4W
  = \int_W G^{\alpha \beta} (\nabla_\beta v_\alpha) \; d^4W
\label{eq:GDiv}
  \end{equation}
 Over small enough volumes this, together with the Einstein Equations,
gives the local conservation of matter.  It should be noted that
$n^\alpha$ is the normal to the junction surface $\Sigma$, and
$m_\alpha$ is the normal to $S$, the closed boundary of $W$; the two
are quite different in general and, even if the two surfaces coincide
partially, they may still differ in sign there.

     Consider now a spacelike boundary surface $\Sigma$, where no
signature change occurs, that divides $W$ and $S$ into two parts $W_+$
\& $W_-$, $S_+$ \& $S_-$, $S_o = \Sigma \cap W$ being the enclosed
region of $\Sigma$, as in fig 1.  In general this ${\bf p}$ (given by
(\ref{eq:p}) and (\ref{eq:pG})) is not even C$^0$ through $\Sigma$.
However the Divergence Theorem holds within each part, so adding them
gives
 \begin{eqnarray}
    & & \int_{S_+} G^{\alpha\beta}_+ v_\alpha^+ m_\beta^+ \; d^3S
 + \int_{S_o} G^{\alpha\beta}_+ v_\alpha^+ (+n_\beta^+) \; d^3S
\nonumber  \\
 & + & \int_{S_-} G^{\alpha\beta}_- v_\alpha^- m_\beta^- \; d^3S
 + \int_{S_o} G^{\alpha\beta}_- v_\alpha^- (-n_\beta^-) \; d^3S
\nonumber  \\
    & = & \int_{W_+} \nabla_\beta^+ (G^{\alpha\beta}_+ v_\alpha^+) \;
d^4W
    + \int_{W_-} \nabla_\beta^- (G^{\alpha\beta}_- v_\alpha^-) \; d^4W
     \end{eqnarray}
 \begin{equation}  \Rightarrow \oint_{S} G^{\alpha\beta} v_\alpha
m_\beta \; d^3S
     + \int_{S_o} \left[ G_{\alpha\beta} v^\alpha n^\beta \right] \;
d^3S
  = \int_{W} \nabla_\beta (G^{\alpha\beta} v_\alpha) \; d^4W
\label{eq:GDivJ1}
 \end{equation}
 where $d^3S$ and $d^4W$ can be made smooth well defined volume
elements, by a suitable choice of coordinates spanning $\Sigma$,
provided only (\ref{eq:Dg}) is satisfied.
 \vspace*{5cm}      %(FIG 1 HERE)}
 \begin{center}
 \parbox{14cm}{ \small
{\bf Fig 1.}~~~~~(a) The volume $W$ with boundary $S$, which
intersects a spacelike surface of discontinuity $\Sigma$, where the
manifold is only $C^2$.  Since a manifold must be $C^4$ (i.e.
$g_{\alpha \beta}$ must be $C^3$) to satisfy the conservation
equations $\nabla_\mu G^{\mu \nu} = 0$, we wish to determine whether
they hold across $\Sigma$. ~~~~~(b) The boundary $\Sigma$ divides $W$
and $S$ into $W_+$, $W_-$, $S_+$ and $S_-$, and the enclosed region of
$\Sigma$, $S_o$, completes the boundaries.  Application of the
Divergence Theorem in each part, supplemented by the Darmois-Israel
junction conditions, shows that matter is conserved if the signature
stays Lorentzian, even though not all components of $G^{\mu \nu}$ are
continuous.  Similar results hold for the electromagnetic and other
fields. We write $m^\alpha$ for the normal to the volume boundary used
in the Divergence theorem, and $n^\alpha$ for the normal to the
surface of discontinuity $\Sigma$ used in the junction conditions.
The two normals to $S_o$ are shown on both sides of $\Sigma$.
Unlabelled vectors are $m^\alpha$ on $S_+$ and $S_-$.
 }
 \end{center}
 \noindent We now choose $v^\gamma$ to be each of the basis vectors of
normal coordinates $n^\gamma$ and $e_i^\gamma$ in turn, and insert the
Israel identities (\ref{eq:Is1}) \& (\ref{eq:Is2}) in
(\ref{eq:GDivJ1}) to obtain the appearance of the Divergence Theorem
for $G^{\sigma \delta} v_\sigma$ (\ref{eq:GDiv}) as if no
discontinuity were present.  (Note that the volume integrands can be
evaluated to
 \begin{eqnarray}
    \nabla_\beta (G_\alpha^\beta e_i^\alpha) = G_\alpha^\beta
(\nabla_\beta e_i^\alpha)
            & = & \tilde{G}^j_k \tilde{\Gamma}^k_{ij}  \\
         \nabla_\beta (G_\alpha^\beta n^\alpha) = G_\alpha^\beta
(\nabla_\beta n^\alpha)
            & = & \tilde{G}^j_k K^k_j  \end{eqnarray}
 and since $\tilde{G}_{ij} = G_{\alpha\beta}\:e_i^\alpha\:e_j^\beta$
contains $\tilde{\partial}_o K_{ij} = \frac{1}{2} \tilde{g}_{ij,o,o}$
 --- see eq (\ref{eq:Gee}) --- they are not continuous across
$\Sigma$, and $\Psi^\delta$ is not $C^1$.)  Thus D provide the
necessary link that ensures conservation of matter across boundary
surfaces.

     For the case when $\Sigma$ is a surface layer, only the first
fundamental form is continuous, so we must impose conservation in some
other way, as was done in [22].  (See also [23].)  We might then
rewrite (\ref{eq:GDivJ1}) to include the surface layer in the volume
integrals, using a Dirac delta,
 \begin{equation}  \oint_{S} G^{\alpha\beta} v_\alpha m_\beta \; d^3S
     = \int_{W} \left( \nabla_\beta (G^{\alpha\beta} v_\alpha)
     - \delta(\tau) \left[ G_{\alpha\beta} v^\alpha n^\beta \right]
\right)
           \; d^4W             \label{eq:GDivJ2}  \end{equation}
 or more generally we write
 \begin{eqnarray}  \oint_{S} \Psi^\beta m_\beta \; d^3S
     + \int_{S_o} \left[ \Psi_\beta n^\beta \right] \; d^3S
  & = & \int_{W} \nabla_\beta \Psi^\beta \; d^4W  \label{eq:PDivJ1}
\\
       \oint_{S} \Psi^\beta m_\beta \; d^3S
     & = & \int_{W} \left( \nabla_\beta \Psi^\beta
     - \delta(\tau) \left[ \Psi_\beta n^\beta \right] \right) \; d^4W
                    \label{eq:PDivJ2}  \end{eqnarray}
 We can think of this double application of (\ref{eq:Div}) in a
discontinuous setting as constituting a `patchwork Divergence
theorem'.  \\[8mm]

 \noindent {\bf MODIFYING ISRAEL'S IDENTITIES}

     We now turn to the case of a boundary where a change of signature
occurs.  We continue to use D for matching the gravitational field
across a signature change, since they ensure the geometries of the two
manifolds can fit together at $\Sigma$, and it turns out they require
no modification despite any metric discontinuity in $\tilde{g}_{oo}$.
The first condition ensures the induced metric on $\Sigma$ is the same
from either side, and allows the two truncated manifolds to fit over
the whole surface.  The second ensures continuity of affine structure,
as indicated by (\ref{eq:ncK}).  The metric is clearly less continuous
than in the L to L case, and the Lichnerowicz conditions are no longer
equivalent, since one can no longer find admissible coordinates in
which the full
 4-metric is continuous and non-degenerate through $\Sigma$.

     We now follow Israel's procedure very closely.  He defines the
normal to $\Sigma$ by
 \begin{equation}  n^\alpha n_\alpha = \epsilon = \pm 1 \end{equation}
 such that $\epsilon = +1$ (or $-1$) for a spacelike (or timelike)
normal (timelike or spacelike $\Sigma$) respectively.  Of course,
$\epsilon$ does not change across $\Sigma$ in his case, as there is no
change of signature. For our purposes, we know $\Sigma$ must be
spacelike for a signature change, so instead we set $\epsilon = +1$ on
the E side $M^+$, and $\epsilon = -1$ on the L side $M^-$.

     At this point the existence of two different normal vectors
becomes apparent.  Recall that $\tau = \xi^o$ is the proper
time/distance coordinate of geodesic normal coordinates, as defined
earlier.  The gradient of the function $\tau = \tau(x^\gamma)$ is
 \begin{equation}  l_\beta = \frac{\partial \xi^o}{\partial x^\beta}
        = \overline{e}_\beta^o  \end{equation}
 where $\overline{e}^o_\gamma$ is one of the dual basis vectors of
geodesic normal coordinates, and the tangent vector to the $\tau$
coordinate lines is
 \begin{equation}  n^\beta = \frac{\partial x^\beta}{\partial \xi^o}
      = e_o^\beta  \end{equation}
 The former gives the sense in which $\tau$ increases and the latter
points in the positive $\tau$ direction, i.e. they both `point' into
$M^+$, the E region.  Thus we have
 \begin{equation}
  l_\alpha l^\alpha = \epsilon = n^\alpha n_\alpha  \mbox{~~,~~~~~}
l_\alpha n^\alpha = 1
  \end{equation}
 so that $\tilde{l}_\mu = \delta^o_\mu$ and $\tilde{n}^\nu =
\delta_o^\nu$ are continuous through $\Sigma$, but
 \begin{equation}
  n_\alpha = \epsilon l_\alpha  \mbox{~~~~~and~~~~~}  l^\alpha =
\epsilon n^\alpha  \end{equation}
 so $\tilde{l}^\mu = \epsilon \delta_o^\mu$ and $\tilde{n}_\nu =
\epsilon \delta^o_\nu$ are not.  We will call $l_\alpha$ the
``gradient normal", and $n^\alpha$ the ``tangent normal".  Note that
${\bf l} = l_\alpha {\bf d}^\alpha$ is a
 one-form, and ${\bf n} = n^\alpha \mbox{\boldmath $\partial$}_\alpha$
is a vector.  To establish which of these is the appropriate one to
use in the
 Darmois-Israel matching, we specify that we want the 3-metric
$\tilde{g}_{ij}$ of eq (\ref{eq:ncg}) to be a $C^1$ function of the
normal coordinate $\tau$, which leads to
 \begin{eqnarray}
    0 = [K_{ij}] & = & [\frac{1}{2} \tilde{g}_{ij,o}] \nonumber \\
 & = & [\frac{1}{2}
            \{ (e_o^\gamma \partial_\gamma g_{\alpha \beta})
e_i^\alpha e_j^\beta
     + g_{\alpha \beta} (\tilde{\partial}_o e_i^\alpha) e_j^\beta
   + g_{\alpha \beta} e_i^\alpha (\tilde{\partial}_o e_j^\beta) \}]
\nonumber \\
 & = & [\frac{1}{2} e_i^\alpha e_j^\beta (e_o^\gamma g_{\alpha \beta ,
\gamma}
   + g_{\beta \gamma} \partial_\alpha e_o^\gamma
   + g_{\alpha \gamma} \partial_\beta e_o^\gamma)]  \nonumber \\
 & = & [\frac{1}{2}
          e_i^\alpha e_j^\beta ( - 2 e_o^\gamma \Gamma_{\gamma, \alpha
\beta}
  + \partial_\alpha (g_{\beta \gamma} e_o^\gamma)
  + \partial_\beta (g_{\alpha \gamma} e_o^\gamma) )]  \nonumber \\
 & = & [\frac{1}{2} e_i^\alpha e_j^\beta \epsilon (2 \partial_\alpha
  \overline{e}^o_\beta - 2 \overline{e}^o_\gamma \Gamma^\gamma_{\alpha
\beta})]
         \nonumber \\
 & = & [e_i^\alpha e_j^\beta (\nabla_\alpha n_\beta)]  \end{eqnarray}
 At a surface of signature change, then, the extrinsic curvature that
must be matched across $\Sigma$ is defined relative to $\Sigma$'s
tangent normal
 --- a unit normal vector whose contravariant components point from
$M^-$ to $M^+$ on both sides.

  Apart from some sign mistakes, not all corrected in the errata (see
reference), Israel's working up to his eqs (12)-(15) carries over
without change.  Using $K = K_m^m = {}^3 \!\! g^{ml}K_{ml}$, we have
 \begin{equation} G_{\alpha \beta} \: n^\alpha n^\beta = \frac{1}{2}
   \left\{ K^2 - K_{ij} K^{ij} - \epsilon \: {}^3 \!\! R \right\}
     \label{eq:Gnn}  \end{equation}
 \begin{equation} G_{\alpha \beta} \: n^\alpha e_i^\beta
           = {}^3 \! \nabla_j K_i^j - {}^3 \! \nabla_i K
\label{eq:Gne}
  \end{equation}
 and the remaining components are
 \begin{equation} G_{\alpha \beta} \: e_i^\alpha e_j^\beta
    =  {}^3 \! G_{ij} - \epsilon \: {}^3 \!\! g_{ik} \left\{
        \partial_o K^k_j + K K^k_j
      - \frac{1}{2} \delta^k_j \left( 2 \partial_o K + K^2
  + K_{lm} K^{lm} \right) \right\}   \label{eq:Gee}  \end{equation}
 where ${}^3 \! G_{ij}$, ${}^3 \!\! R$ and ${}^3 \! \nabla_i$ are the
3 dimensional intrinsic Einstein tensor, Ricci scalar and covariant
derivative of $\Sigma$.  (See also [24] but note that their definition
of $K_{ij}$ is the negative of our eq (\ref{eq:Kij}).)  D keep
everything on the {\it rhs} of (\ref{eq:Gnn}) \& (\ref{eq:Gne})
unchanged except for $\epsilon$.  Thus the modified Israel identities
are
 \begin{eqnarray}  [ G_{\alpha \beta} \: n^\alpha n^\beta ] =
[\tilde{G}_{oo}]
                      & = & - {}^3 \!\!R       \label{eq:modIs1}  \\
         {[ G_{\alpha \beta} \: n^\alpha e_i^\beta ]} =
[\tilde{G}_{oi}]
                        & = & 0  \label{eq:modIs2}  \end{eqnarray}
 Since the operation of raising and lowering indices is not smooth
through $\Sigma$, this implies
 \begin{eqnarray} [\tilde{G}^{oo}] = [G^{\alpha \beta} l_\alpha
l_\beta]
                          & = & - {}^3 \!\!R  \label{eq:modIs3}  \\
        {[\tilde{G}^o_o]} = [G^\alpha_\beta l_\alpha n^\beta]
                   & = & K^2 - K_{ij} K^{ij}  \label{eq:modIs4}  \\
        {[\tilde{G}_o^i]} = [ G_\alpha^\beta \: n^\alpha
\overline{e}^i_\beta ]
                                        & = & 0  \label{eq:modIs5}  \\
        {[\tilde{G}^{oi}]} = [G^{\alpha \beta} l_\alpha
\overline{e}^i_\beta]
       & = & 2({}^3 \! \nabla_j K^{ij} - {}^3 \!\! g^{ij} {}^3 \!
\nabla_j K)
                                              \label{eq:modIs6}  \\
        {[\tilde{G}^o_i]} = [G^\alpha_\beta l_\alpha e_i^\beta]
   & = & 2({}^3 \! \nabla_j K_i^j - {}^3 \! \nabla_i K)
\label{eq:modIs7}
 \end{eqnarray}  \\[1cm]

 \noindent {\bf GENERALISING THE PATCHWORK DIVERGENCE THEOREM}

     A change of signature, being a metric phenomenon, should affect
the Divergence Theorem (\ref{eq:Div}), but not Stokes' Theorem
(\ref{eq:Stokes}).  In other words, if $\bf p$ satisfies Stokes'
Theorem on a particular manifold when the signature doesn't change,
then the same $\bf p$ must still satisfy it on the same manifold when
the signature does change.  Since this reasoning is not valid for
$\Psi^\delta$, we must adapt the patchwork approach to the case of
signature change at $\Sigma$.  Consequently we assume that there
exists an orientation (a smooth non-zero form) right through $\Sigma$,
so that Stokes' Theorem holds for a sufficiently smooth $\bf p$.
However we will not actually need to assume that $\bf p$ is $C^1$.
Although $\tilde{g}_{oo} = \epsilon$ is discontinuous, being double
valued on the identified boundary $\Sigma^+ \equiv \Sigma^-$, the
volume element in normal coordinates $d^4\tilde{W} = \sqrt{\epsilon g}
\; \varepsilon_{\alpha \beta \gamma \delta} \; d\xi^\alpha d\xi^\beta
d\xi^\gamma d\xi^\delta$ is actually smooth through $\Sigma$.  This
does not make (\ref{eq:Div}) valid even if $\Psi^\delta$ is smooth
through a signature change, since the conversion of Stokes Theorem to
the Divergence theorem involves the metric itself.

     In order to preserve some kind of `Divergence Theorem' in this
case, we once again use the usual Divergence Theorem on either side of
$\Sigma$ and join them by means of junction conditions on
$\Psi^\alpha$ appropriate for signature change, such as the modified
Israel Identities, which now give a non-zero surface contribution to
the volume integral.  In other words, we expect a result of the form
(\ref{eq:PDivJ1}) or (\ref{eq:PDivJ2})
 \begin{eqnarray}  \oint_{S} \Psi^\beta m_\beta \; d^3S
     - \int_{S_o} E \; d^3S
  & = & \int_{W} \nabla_\beta \Psi^\beta \; d^4W  \label{eq:mPDivE1}
\\
       \oint_{S} \Psi^\beta m_\beta \; d^3S
     & = & \int_{W} \left( \nabla_\beta \Psi^\beta
     + \delta(\tau) E \right) \; d^4W
                    \label{eq:mPDivE2}  \end{eqnarray}
 where $\tau = 0$ on $\Sigma$, which we now derive, obtaining $E$ for
this case.  We point out that the normal vector $m_\alpha$ is really a
gradient normal, that `points' out of $W$.  In other words, if we set
up geodesic normal coordinates based on $S$, with proper time/distance
coordinate $\zeta^4 = 0$ on $S$ and increasing outwards, and
$\zeta^b$, $b = 1,2,3$ surface coordinates, then $m_\alpha =
\partial_\alpha \zeta^4$.  However it is customary to define the
direction of $m_\alpha$ via the pointing behaviour of its metric dual
$m^\alpha$, given after eq (\ref{eq:Div}).  For an arbitrary volume
$W$ bounded by $S$ and spanning $\Sigma$, we again split $W$ as in fig
2.
 \vspace*{5cm}     % (FIG 2 HERE)
 \begin{center}
 \parbox{14cm}{ \small
{\bf Fig 2.}~~~~~In the case when the signature changes from $(-+++)$
in $W_-$ to $(++++)$ in $W_+$, $m^\alpha_+$ changes direction.
 }
 \end{center}

     The usual Divergence Theorem applies to the volume integrals over
$W_+$ and $W_-$, hence, noting that the gradient normals,
$m_\alpha^\pm$, on $S_o^+$ and $S_o^-$ convert into the gradient
normals $-l_\alpha^+$ and $+l_\alpha^-$ on $\Sigma^+$ and $\Sigma^-$
respectively, and being careful not to change index positions, we find
 \[    \int_{S_+} \Psi^\alpha_+ m_\alpha^+ \; d^3S
       + \int_{S_o} \Psi^\alpha_+ (-l_\alpha^+) \; d^3S
       + \int_{S_-} \Psi^\alpha_- m_\alpha^- \; d^3S
       + \int_{S_o} \Psi^\alpha_- (+l_\alpha^-) \; d^3S    \]
 \begin{equation}  = \int_{S_+} \Psi^\alpha_+ m_\alpha^+ \; d^3S
       + \int_{S_-} \Psi^\alpha_- m_\alpha^- \; d^3S
       - \int_{S_o} [\Psi^\alpha l_\alpha] \; d^3S
       = \int_{W_+} \nabla_\alpha^+ \Psi^\alpha_+ \; d^4W
       + \int_{W_-} \nabla_\alpha^- \Psi^\alpha_- \; d^4W
               \label{eq:mPDivJ0}  \end{equation}
 which holds for arbitrary $W$ and $S_o$, so we conclude that
 \begin{equation}  E = ( \Psi^\alpha_+ l_\alpha^+ - \Psi^\alpha_-
l_\alpha^- )
       = [\Psi^\alpha l_\alpha]   \label{eq:fjump}  \end{equation}
 We can represent this in the following two forms, which allow easier
comparison with (\ref{eq:Div}) and (\ref{eq:PDivJ1})-(\ref{eq:PDivJ2})
 \begin{eqnarray}   \Rightarrow \oint_{S} \Psi^\alpha m_\alpha \; d^3S
       - \int_{S_o} [ \Psi^\alpha l_\alpha ] \; d^3S
        & = & \int_{W} \nabla_\alpha \Psi^\alpha \; d^4W
\label{eq:mPDivJ1}  \\
     \oint_{S} \Psi^\alpha m_\alpha \; d^3S & = & \int_{W}
(\nabla_\alpha \Psi^\alpha
   + \delta(\tau)[\Psi^\alpha l_\alpha]) \; d^4W  \label{eq:mPDivJ2}
   \end{eqnarray}
 These forms may be justified on the grounds that $d^4\tilde{W}$ and
$d^3\tilde{S}$ are smooth through $\Sigma$, and no further
manipulation with a discontinuous $\tilde{g}_{\mu \nu}$ is required.
(However they are not well defined if the coordinates near $\Sigma$
are {\it defined} to be such that $g_{oo} \rightarrow 0$ on $\Sigma$.
Eq (\ref{eq:mPDivJ0}) is always well defined.)  In contrast to the
case of no signature change, where the substitution $[\Psi^\alpha
l_\alpha] = -[\Psi_\alpha n^\alpha]$ does not affect the validity of
eqs (\ref{eq:PDivJ1})-(\ref{eq:PDivJ2}), it is important to use only
the gradient normal here.  If we know how $\Psi^\alpha$ matches across
$\Sigma$ we can use this to determine the surface ``singularity" on
$\Sigma$ associated with $\Psi^\alpha$ due to the signature change.
The surface term {\it only} disappears for smooth contravariant
$\Psi^\delta$ {\it in normal coordinates}.  (Recall that the Euclidean
region is ``+", the Lorentzian region is
 ``--", and $l_\alpha^\pm$ `point' into the Euclidean region.)
Results (\ref{eq:mPDivJ0})-(\ref{eq:mPDivJ2}) are of course valid
whether or not the signature changes, and for timelike or spacelike
$\Sigma$, in all viable combinations, whereas (\ref{eq:PDivJ1})-
(\ref{eq:PDivJ2}) are only valid for constant Lorentzian signature at
a spacelike $\Sigma$. \\[1cm]

 \noindent {\bf NON-CONSERVATION OF MATTER?}

     Returning to the construction of fig 2 with $\Psi^\beta =
G^{\alpha \beta} v_\alpha$, it is clear that there must be a source
term $E(v_\gamma)$ on $\Sigma$ in the volume integral over
$\nabla_\beta (G^{\alpha \beta} v_\alpha)$, which depends on the
choice of a {\it covariant} $v_\gamma$.  This choice of $\Psi^\beta$
in (\ref{eq:fjump}) gives
 \begin{equation}  E(v_\gamma) =  G^{\alpha\beta}_+ v_\alpha^+
l_\beta^+ \: - \:
  G^{\alpha\beta}_- v_\alpha^- l_\beta^- = [G^{\alpha\beta} v_\alpha
l_\beta]
                                      \label{eq:GEv}  \end{equation}
 For $v_\alpha = l_\alpha$ and $\overline{e}^i_\alpha$ we find
respectively
 \begin{eqnarray}  E(l_\alpha) & = & [\tilde{G}^{oo}]
                                 = - {}^3 \!\!R  \label{eq:Egrav0}  \\
  E(\overline{e}^i_\alpha) & = & [\tilde{G}^{oi}] = 2({}^3 \!
\nabla_j K^{ij}
         - {}^3 \!\! g^{ij} {}^3 \! \nabla_j K)  \label{eq:Egravi}
   \end{eqnarray}
 If instead of (\ref{eq:pG}) we choose $\Psi^\beta = G_\alpha^\beta
v^\alpha$ with smooth {\it contravariant} $v^\alpha$, then $E$ depends
on $v^\alpha$ and we arrive at
 \begin{eqnarray}  E(n^\alpha) & = & [\tilde{G}_o^o]
                         = K^2 - K_{ij} K^{ij}  \label{eq:Egrav0b}  \\
  E(e_i^\alpha) & = & [\tilde{G}^o_i] =
    2({}^3 \! \nabla_j K_i^j - {}^3 \! \nabla_i K)  \label{eq:Egravib}
  \end{eqnarray}
 $\tilde{G}^{oo}_-$ and $\tilde{G}^{io}_-$ are the energy density and
the 3 energy fluxes/momentum densities on the L side, as measured by
an observer moving orthogonally to the transition surface.  The
meanings of the Euclidean quantities $\tilde{G}^{oo}_+$ and
$\tilde{G}^{io}_+$ are open for discussion. \\[1cm]

 \noindent {\bf OTHER MATCHING OPTIONS}

     The results presented above are based on those junction
conditions that we regard as the most reasonable.  For the sake of
completeness we mention two important steps at which a different
choice of sign has a large effect on the results.

     The first one is in the definition of the extrinsic curvature.
If instead of (\ref{eq:Kij}), we choose
 \begin{equation}
   K'_{ij} = (\nabla_\beta l_\alpha) \: e_i^\alpha \, e_j^\beta
\end{equation}
 then we find
 \begin{equation}
   [K'_{ij}] = 0  \Rightarrow  K_{ij}^+ = - K_{ij}^-  \end{equation}
 and the modified Israel identities become
 \begin{equation}  [ G_{\alpha \beta} \: n^\alpha n^\beta ] =
[\tilde{G}_{oo}]
                                      = - {}^3 \!\!R  \end{equation}
 \begin{equation}  [ G_{\alpha \beta} \: n^\alpha e_i^\beta ] =
[\tilde{G}_{oi}]
         = 2({}^3 \! \nabla^j K_{ij}^+ - {}^3 \! \nabla_i K^+)
\end{equation}
 so, giving the $K_{ij}$ their values in $M^+$, the {\it rh} sides of
eqs (\ref{eq:modIs2}) and (\ref{eq:modIs7}) are swopped and the {\it
rh} sides of eqs (\ref{eq:modIs5}) and (\ref{eq:modIs6}) are swopped
if this sign is changed.  This does mean that
$E(\overline{e}^i_\alpha) = 0$, but $E(l_\alpha)$ and $E(n^\alpha)$
are unchanged.  Also $[K'_{ij}] = 0$ implies $\tilde{g}^+_{ij,o} = -
\tilde{g}^-_{ij,o}$.

     The second sign choice relates the orientations of the manifolds
$M^-$ and $M^+$.  If we don't assume that the combined manifold is
oriented, then there are two possible relative orientations; one gives
eq (\ref{eq:GEv}) and the other leads to
 \begin{equation}
  E(v_\gamma) =  -(G^{\alpha\beta}_+ v_\alpha^+ l_\beta^+ \: + \:
                 G^{\alpha\beta}_- v_\alpha^- l_\beta^-)
\end{equation}

     Changing this sign factor swops the {\it rh} sides of eqs
(\ref{eq:modIs1})/(\ref{eq:modIs3}), (\ref{eq:modIs2}), and
(\ref{eq:modIs5}) with (\ref{eq:modIs4}), (\ref{eq:modIs7}), and
(\ref{eq:modIs6}) respectively, and changes all their signs.  There
are no combinations of choices which will make the surface terms
$E(l_\alpha)$ or $E(n^\alpha)$ disappear, because eq (\ref{eq:Gnn}) is
second order in the $K_{ij}$ and some but not all of its terms contain
$\epsilon$.  \\[1cm]

 \noindent {\bf ELECTROMAGNETIC FIELDS AND SOURCES}

     We here take the electromagnetic field as an example of a vector
field, and we assume either vacuum or no change in dielectric
properties at the transition surface.  According to [25] the junction
conditions for a macroscopic electromagnetic (EM) field at a
dielectric boundary
 --- which for his purposes is actually timelike (i.e. no signature
change, L-L) --- are:
 \begin{equation}         {[ {\bf D}_\perp ]} = 0 \mbox{~~,~~~~~}
              {[ {\bf E}_\parallel ]} = 0 \mbox{~~,~~~~~}
              {[ {\bf B}_\perp ]} = 0 \mbox{~~,~~~~~}
              {[ {\bf H}_\parallel ]} = 0     \end{equation}
 which means that for the microscopic quantities, vectors ${\bf E}$
and ${\bf B}$ are C$^0$.  We now try to find juction conditions for
the EM potential which are analogous to D for the gravitational
potentials.  Working temporarily in normal coordinates at a timelike
or spacelike boundary, and assuming that D are satisfied, this can be
ensured by, (i) choosing a ``normal gauge" (the equivalent of normal
coordinates),
 \begin{equation}  \tilde{A}_o^\Sigma = 0  \end{equation}
 on the boundary, which ensures $\tilde{\partial}_i \tilde{A}_o$,
$\tilde{\partial}_j \tilde{\partial}_i \tilde{A}_o$ are all zero, (ii)
then requiring
 \begin{equation}
  [\tilde{A}_i] = 0 \mbox{~~~~~and~~~~~} [\tilde{\partial}_o
\tilde{A}_j] = 0  \end{equation}
 everywhere on the surface, which of course means $[\tilde{\partial}_i
\tilde{\partial}_o \tilde{A}_j]$, $[\tilde{\partial}_i \tilde{A}_j]$
and $[\tilde{\partial}_k \tilde{\partial}_i \tilde{A}_j]$ are also
zero.  These compare nicely with D in normal coords: $[\tilde{g}_{ij}]
= 0 = [\tilde{\partial}_o \tilde{g}_{ij}]$.  In non-normal coords, the
gauge and junction conditions are
 \begin{equation}  A_\alpha^\Sigma n^\alpha = 0 \mbox{~~,~~~~~}
[A_\alpha e_i^\alpha] = 0 \mbox{~~~~~and~~~~~}
           [n^\alpha \partial_\alpha (A_\beta e_i^\beta)] = 0
\label{eq:Ajc1}
  \end{equation}
 and if $[K_{ij}] = 0$ these can be written as
 \begin{equation}  A_\alpha^\Sigma n^\alpha = 0 \mbox{~~,~~~~~}
[A_\alpha e_i^\alpha] = 0 \mbox{~~~~~and~~~~~}
             [e_i^\beta n^\alpha \nabla_\alpha A_\beta] = 0
\label{eq:Ajc2}
  \end{equation}
 In terms of $\tilde{F}_{\mu \nu}$ we get
 \begin{equation}  [\tilde{F}_{oi}] = 0 \mbox{~~,~~~~~}
[\tilde{F}_{ij}] = 0 \mbox{~~,~~~~~}
  [\tilde{\nabla}_k \tilde{F}_{oi}] = 0 \mbox{~~,~~~~~}
[\tilde{\nabla}_k \tilde{F}_{ij}] = 0
                                     \label{eq:Fjc}  \end{equation}
 and, for a timelike boundary, Jackson's conditions are recovered.
Because these quantities are coordinate invariant and projected onto
the boundary surface, they are unaffected by a change in signature.
However, non-dummy indices may not be raised and lowered freely, and
of course (\ref{eq:Ajc1}) and (\ref{eq:Ajc2}) aren't gauge invariant.

     Now the current density
 \begin{equation}  4 \pi J_\beta = \nabla^\alpha F_{\alpha \beta} =
\nabla^\alpha
   \partial_\alpha A_\beta - \nabla^\alpha \partial_\beta A_\alpha
\end{equation}
 and the stress-energy tensor
 \begin{equation} 4 \pi T_{\alpha \beta} = F_\alpha^\mu F_{\beta \mu}
   - \frac{1}{4} g_{\alpha \beta} F^{\mu \nu} F_{\mu \nu}
\end{equation}
 have normal gauge, normal coordinate components
 \begin{eqnarray}  4 \pi \tilde{J}_j & = &
        \tilde{g}^{im} \tilde{\nabla}_i ( \tilde{\partial}_m
\tilde{A}_j
   - \tilde{\partial}_j \tilde{A}_m ) + \epsilon \tilde{\nabla}_o (
\tilde{\partial}_o
     \tilde{A}_j - \tilde{\partial}_j \tilde{A}_o )  \label{eq:4pJi}
\\
    & = & {}^3 \! \tilde{\nabla}^m \tilde{F}_{mj} + \epsilon (
\tilde{\partial}_o
    \tilde{F}_{oj} + K \tilde{F}_{oj} - 2 K^m_j \tilde{F}_{om} )  \\
      4 \pi \tilde{J}_o & = & \tilde{g}^{im} \tilde{\nabla}_i (
\tilde{\nabla}_m \tilde{A}_o
    - \tilde{\nabla}_o \tilde{A}_m ) = {}^3 \! \nabla^m \tilde{F}_{mo}
\label{eq:4pJo}
 \end{eqnarray}
 (where the terms containing $\epsilon$ in $\tilde{J}_o$ cancel owing
to the antisymmetry of $F_{\alpha \beta}$) and
 \begin{eqnarray} 4 \pi \tilde{T}_{oo} & = &
         \frac{1}{2} \tilde{F}_{ok} \tilde{F}_{ol} g^{kl}
       - \frac{1}{4} \epsilon \tilde{F}^{kl} \tilde{F}_{kl}  \\
  4 \pi \tilde{T}_{oi} & = & \tilde{F}_{ok} \tilde{F}_{il}
\tilde{g}^{kl}  \\
  4 \pi \tilde{T}_{ij} & = & \epsilon ( \tilde{F}_{oi} \tilde{F}_{oj}
    - \frac{1}{2} \tilde{g}_{ij} \tilde{F}_{ok} \tilde{F}_{ol}
\tilde{g}^{kl} )
    + ( \tilde{F}_{ik} \tilde{F}_{jl} \tilde{g}^{kl}
    - \frac{1}{4} \tilde{g}_{ij} \tilde{F}_{kl} \tilde{F}_{mn}
            \tilde{g}^{km} \tilde{g}^{ln} ) \end{eqnarray}
 With the foregoing EM junction conditions and the standard D
conditions, we find
 \begin{eqnarray}
    {4 \pi [\tilde{J}_j]} = {4 \pi [J_\alpha e_j^\alpha]} & = &
   \epsilon^+ \tilde{\partial}_o^+ \tilde{F}_{oj}^+ - \epsilon^-
     \tilde{\partial}_o^- \tilde{F}_{oj}^- + (\epsilon^+ - \epsilon^-)
      (K \tilde{F}_{oj} - 2 K^m_j \tilde{F}_{om}) \\
 {4 \pi [\tilde{J}_o]} = {4 \pi [J_\alpha n^\alpha]} & = & 0
                 \label{eq:J0jump}  \\
       4 \pi [\tilde{J}^o] = 4 \pi [J^\alpha l_\alpha] & = &
    (\epsilon^+ - \epsilon^-) {}^3 \! \tilde{\nabla}^m \tilde{F}_{mo}
 \end{eqnarray}
 (c.f. [26].)  Across a L-L boundary, $\epsilon^+ = \epsilon^-$, the
entire stress energy tensor is continuous, while $[J^\alpha l_\alpha]
= 0$ links the Divergence Theorems on either side of $\Sigma$ and
ensures conservation of
 4-current.  However, a jump in the value of the current parallel to
the surface is quite acceptable.

     Returning to eq (\ref{eq:fjump}), for the surface term $E$ in the
case of a signature change, we find that
 \begin{equation}
  E = [J^\alpha l_\alpha] = [\tilde{J}^o] = {}^3 \! \tilde{\nabla}^m
(\tilde{F}_{mo}^+
  + \tilde{F}_{mo}^-) = 2 {\bf \tilde{\mbox{\boldmath $\nabla$}} \cdot
\tilde{E}}
           \label{eq:Eem}  \end{equation}
 where $\tilde{J}^o_-$ is the charge density and $\tilde{F}_{mo}^-$ is
the electric field, both as measured by an orthogonally moving
observer on the L side.  Furthermore, from
 \begin{eqnarray}   4 \pi [\tilde{T}_{oo}] = 4 \pi [\tilde{T}^{oo}] &
= &
      - \frac{1}{4} ( \tilde{F}^{kl}_+ \tilde{F}_{kl}^+
      + \tilde{F}^{kl}_- \tilde{F}_{kl}^- ) = - \frac{1}{2}{\bf
\tilde{B}}^2
                                             \label{eq:EMTjump1}  \\
     4 \pi [\tilde{T}^o_i] = 4 \pi g_{ij} [\tilde{T}^{oj}] & = &
    \tilde{g}^{kl} ( \tilde{F}_{ok}^+ \tilde{F}_{il}^+
    + \tilde{F}_{ok}^- \tilde{F}_{il}^-) = 2 ({\bf \tilde{B}
          \times \tilde{E}})_i  \\
      4 \pi [\tilde{T}^o_o] & = & \frac{1}{2} \tilde{g}^{kl} (
\tilde{F}_{ok}^+
    \tilde{F}_{ol}^+ + \tilde{F}_{ok}^- \tilde{F}_{ol}^- ) = {\bf
\tilde{E}}^2  \\
          4 \pi \tilde{g}_{ij} [\tilde{T}^j_o] = 4 \pi
[\tilde{T}_{oi}] & = & 0
                                   \label{eq:EMTjump4}  \end{eqnarray}
 we see that the EM energy density $\tilde{T}^{oo}$ cannot be
continuous, unless the magnetic field $\tilde{F}_{ij}$ is zero.  This
is also true for $\tilde{T}_{oo}$, whereas $\tilde{T}_o^o$ can only be
continuous if the electric field $\tilde{F}_{oi}$ is zero, and the
continuity of $\tilde{T}^o_i$ and $\tilde{T}^{oi}$ requires a zero
Poynting vector.  If one requires continuity of all components of the
EM stress tensor in all index positions, the entire EM field must be
zero at a signature change, regardless of sign choices in the matching
conditions.  In fact, zero field is required just to make
$\tilde{T}_{oo}$ and $\tilde{T}_o^o$ continuous, so, although we could
recover conservation of 4-current by matching $\tilde{F}^{\mu
\nu}_\pm$ instead of $\tilde{F}_{\mu \nu}^\pm$, there is no way to
recover full EM energy conservation without restricting the field
configuration at the transition surface.  For an electrovac model, the
Einstein equations plus
 (\ref{eq:Egrav0})-(\ref{eq:Egravib}) and
 (\ref{eq:EMTjump1})-(\ref{eq:EMTjump4}) lead to
 \begin{eqnarray}
   {}^3 \!\! R & = & {\bf \tilde{B}}^2  \label{eq:EV1}  \\
   2 ( {}^3 \! \nabla_j K^j_i - {}^3 \! \nabla_i K ) & = &
        4 ( {\bf \tilde{B} \times \tilde{E}} )_i \\
        K^2 - K_{ij} K^{ij} & = & 2 {\bf \tilde{E}}^2
                               \label{eq:EV3}  \end{eqnarray}
 which restrict both the extrinsic curvature of $\Sigma$ and the EM
field configuration.

        We managed to sidestep the question of whether to match
$\tilde{A}^o_\pm$ or $\tilde{A}_o^\pm$ by choosing $\tilde{A}_o^\Sigma
= 0$, but it definitely seems more natural to match
$\tilde{\partial}_o \tilde{A}_i \left. \right|_\pm$ and hence
$\tilde{F}_{oi}^\pm$ than $\tilde{\partial}^o \tilde{A}_i \left.
\right|_\pm$ and $\tilde{F}^{oi}_\pm$.  \\[1cm]

 \noindent {\bf JUNCTION CONDITIONS AND CONSERVATION LAWS}

    In this paper we have considered the effect of signature change on
conservation laws in classical General Relativity.  Though we have
advocated the use of Darmois junction conditions, we emphasise that
results
 (\ref{eq:mPDivJ0})-(\ref{eq:mPDivJ2}),
 (\ref{eq:GEv}), (\ref{eq:Ajc1}),
 (\ref{eq:4pJi})-(\ref{eq:4pJo}) hold for any set of junction
conditions that impose at least (\ref{eq:Dg}) and any set of
coordinates near $\Sigma$ for which the limiting values $\left.
\Psi^\delta l_\delta \right|_\Sigma$ may be calculated, so the surface
effects for any other choice may be calculated this way.
Nevertheless, we are not aware of any junction conditions for
signature change that are less restrictive than D.

     At a change of the signature of spacetime the Darmois conditions
no longer ensure standard matter conservation, as they do through a
constant signature boundary, and only if eqs
 (\ref{eq:Egrav0})-(\ref{eq:Egravib}) are zero can we have matter
conservation in the usual sense, but this means the transition occurs
on a surface with zero extrinsic curvature that has ${}^3 \! R = 0$
 --- a highly restrictive condition, eliminating all realistic
cosmological models.  (Though the $k = 0$ FLRW model survives, its
perturbed cousin does not.)

     Significantly, the condition that the extrinsic curvature be zero
at the signature change surface, $K^\Sigma_{ij} = 0$, which is
required in the Quantum Cosmology approach, does not entirely remove
the need to modify the matter conservation law, since only three of
(\ref{eq:Egrav0})-(\ref{eq:Egravib}) become zero.  If however, one is
satisfied with only $G^\beta_\alpha v^\alpha$ and not $G^{\beta
\alpha} v_\alpha$ being conserved, then $K^\Sigma_{ij} = 0$ will
suffice.

    Since the operation of raising or lowering indices introduces
minus signs on one side of a signature change and not on the other, it
is clear that a non-empty model cannot have all of $G_{\alpha \beta}$
continuous across a signature change as well as all of $G^{\alpha
\beta}$.  If $G_{oi}$ is continuous, then $G^{oi}$ has a jump; if
$G_{oo}$ is continuous then $G^o_o$ has a jump, and vice versa.  The
same applies to any non-zero tensor
 --- if $g_{\alpha \beta}$, $\Gamma_{\alpha, \beta \gamma}$,
$R_{\alpha \beta \gamma \delta}$, $F_{\alpha \beta}$, etc. are
continuous, then $g^{\alpha \beta}$, $\Gamma^\alpha_{\beta \gamma}$,
$R^{\alpha \beta \gamma \delta}$, $F^{\alpha \beta}$, etc. are not.
Thus we cannot expect both $\nabla_\alpha G^{\alpha \beta}$ and
$\nabla_\alpha G^\alpha_\beta$ to be free of surface effects.

     The jump in the value of the energy density relative to the
surface, given by eq (\ref{eq:modIs1}), cannot be viewed as a surface
layer of matter on $\Sigma$ in the usual sense, as Israel's definition
for the surface stress tensor is still zero here.  Rather it is an
effect due to the change in physical law from Lorentzian to Euclidean
forms.  Momentum changes from mass times velocity into the ``momentum"
conjugate to the ``time" coordinate, which is now a spatial direction,
and energy converts to a quantity having the dimensions of
``momentum".

     For the other matching options too, the situation is not much
different.  Matching $\tilde{g}_{ij,o}$ to its negative across the
signature change does remove two out of four surface effects.  But if
orientation is not preserved through the signature change, the
conservation of matter may become separate from the continuity of the
projected Einstein tensor, with conservation breaking down even if all
10 components of $\tilde{G}^{\mu \nu}$ are continuous through
$\Sigma$.

     For any vector field on spacetime, the generalised patchwork
Divergence Theorem, (\ref{eq:mPDivJ1}) or (\ref{eq:mPDivJ2}), and the
resulting expression for $E$, eq (\ref{eq:fjump}), combined with the
appropriate matching conditions, then show that the field may also
have a surface effect due to the signature change and the modification
of its physics.  Specifically, the matching conditions for the EM
field (\ref{eq:Ajc2}) lead to a surface effect in the associated
current density (\ref{eq:Eem}) which is zero only if $\tilde{J}_o =
J_\alpha n^\alpha = 0$.  This implies the current is conserved only if
the net charge density (in geodesic normal coords) is zero everywhere
on the signature change surface, the most likely example being that of
a source-free field.  (It is utterly improbable that a system of
charges and currents should have zero charge density everywhere on a
spacelike slice of the universe.)  However, the EM energy density is
not continuous.  This may not be a problem if there are other fields
or matter components that can exchange energy and momentum with the EM
field, so as to satisfy the overall `modified conservation law'.  But,
for an electrovac solution, (\ref{eq:EV1})-(\ref{eq:EV3}) place very
strong restrictions on the allowed Gravitational and EM fields, which
eliminate any radiation at the transition.  If there is a
 non-zero charge density of electrons, say, then the surface effect
calculated from the EM junction conditions must be consistent with the
results from the Dirac equation, with suitable junction conditions.
This would be an interesting avenue of investigation.

     When considering matter tensors consisting of several components
and/or fields, the above results indicate that we may specify
continuity of momentum density (parallel to $\Sigma$) separately for
each component, but we should expect the individual energy densities
to jump.

     Faced with the inapplicability of the standard Divergence Theorem
across a signature change, our intuition that matter must always be
conserved in the usual way no longer seems physically justified.  In
fact, when considering the physics of signature changes, all intuition
should be very carefully cross-checked.  If we wish to impose extra
restrictions in order to describe a particular physical effect or
situation, we should review the physical justification in the light of
the change in the relevant physical laws.

     At a L to L boundary, D are sufficient to ensure the minimum
necessary continuity and the conservation of all fundamental {\it
gravitational} quantities, and any further restrictions then
specialise to particular scenarios, and eliminate other possibilities.
We have presented a L to E boundary in the same light.  D impose the
same number of conditions, and still ensure minimal continuity and
modified conservation laws, so further restrictions need only be
imposed in order to describe specific physical effects which require
them.  For example, [9] and [10] proposed criteria for when the
signature should change, but these are not specifically required by
the Darmois junction conditions.  Their condition amounted to
requiring continuity of the equation of state (continuous Friedmann
equation) across $\Sigma$.  One might regard this as the equivalent
for a fluid of junction conditions for a field.  Such extra conditions
may often be reasonable and necessary.

     On the other hand, a recent investigation [27,28] related to
Smolin's [29] idea that universes evolve in Darwinian fashion,
required the Darmois approach.  This enabled collapse to a black hole
to pass through a double signature change, emerging into a new
universe.  Interesting results are not possible if $K_{ij} = 0$.

     In the absence of convincing physical arguments or experimental
evidence, as mentioned at the beginning, the ``correct" way to effect
a change of signature remains a matter of conjecture.  The
relationship between the present results and other approaches found in
the literature will be discussed elsewhere. \\[6mm]

 \noindent {\bf ACKNOWLEDGEMENT}

     We wish to thank a referee for pointing out some significant
errors.  TD was partially funded by NSF grant PHY 92-08494.  CH would
like to thank the FRD for a research grant. \\[6mm]

 \setlength{\parindent}{0em}
 {\bf REFERENCES}
 {\small

[1] Hawking S.W. 1982, in {\it Astrophysical Cosmology}, Eds.
Br\"{u}ck Coyne and Longair (Pontifica Academia Scientarium, Vatican
City).

[2] Hartle J.B. and Hawking S.W. 1983, {\it Phys. Rev. D} {\bf 28},
2960-75.

[3] Hawking S.W. 1984, {\it Nucl. Phys.} {\bf B239}, 257-76.

[4] Sakharov, A.D. 1984, {\it Zh. Eksp. Teor. Fiz.} {\bf 87}, 375, and
{\it Sov. Phys. JETP} {\bf 60}, 214-8.

[5] Halliwell J.J. and Hartle J.B. 1990, {\it Phys. Rev. D} {\bf 41},
1815-34.

[6] Gibbons G.W. and Hartle J.B. 1990 {\it Phys. Rev. D} {\bf 42}
2458-68.

[7] Dray, T. Manogue, C.A. Tucker, R.W. 1991, {\it ``Particle
Production from Signature Change", Gen. Rel. Grav.} {\bf 23}, 967-71.

[8] Hayward, S.A. 1992, {\it Class. Q. Grav.} {\bf 9}, 1851.

[9] Ellis, G.F.R. Sumeruk, A. Coule, D. and Hellaby, C. 1992, {\it
``Change of Signature in Classical Relativity", Class. Q. Grav.} {\bf
9}, 1535-54.

[10] Ellis, G.F.R. 1992, {\it ``Covariant Change of Signature in
Classical Relativity", Gen. Rel. Grav.} {\bf 24}, 1047-68.

[11] Dray, T. Manogue, C.A. Tucker, R.W. 1993, In Preparation: {\it
``The Effect of Signature Change on Scalar Field Propagation''},
(Oregon State Univ. preprint).

[12] Hayward, S.A. 1993, {\it Class. Q. Grav} {\bf 10}, L7.

[13] Kerner, R. and Martin, J. 1993, {\it Class. Q. Grav.} {\bf 10},
2111-2122.

[14] Dray, T. Manogue, C.A. Tucker, R.W. 1993, {\it Phys. Rev. D} {\bf
48}, 2587-90.

[15] Darmois, G. 1927, {\it M\'{e}morial des Sciences
Math\'{e}matiques}, (Gauthier-Villars, Paris), {\bf Fasc. XXV}, {\it
``Les Equations de la Gravitation Einsteinienne"}, Chapitre V.

[16] Israel, W. 1966, {\it Nuov. Cim.} {\bf 44B}, 1-14, and
corrections in {\it Ibid} {\bf 48B}, 463.

[17] O'Brien, S. and Synge, J.L. 1952, {\it Communications of the
Dublin Institute for Advanced Studies}, {\bf Series A}, {\bf No 9},
{\it ``Jump Conditions at Discontinuities in General Relativity"}.

[18] Bonnor, W.B. and Vickers, P.A. 1981, {\it Gen. Rel. Grav.} {\bf
13}, 29-36.

[19] Clarke, C.J.S. and Dray, T. 1987, {\it Class. Q. Grav.}, {\bf 4},
265-75.

[20] Lichnerowicz, A. 1955, {\it Th\'{e}ories Relativistes de la
Gravitation et de l'Electromagn\'{e}tisme}, (Masson, Paris), Chapitre
III.

[21] Wald, R.M. 1984, {\it General Relativity}, (University of Chicago
Press), appendix B.

[22] Barrab\`{e}s, C. 1989, {\it Class. Q. Grav.} {\bf 6}, 581-8.

[23] Dray, T. and Padmanabhan, T. 1989, {\it Gen. Rel. Grav.} {\bf
21}, 741.

[24] Lightman, A.P. Press, W.H. Price, R.H. and Teukolsky, S.A. 1975,
{\it Problem Book in Relativity and Gravitation}, (Princeton
University Press), problem 9.32.

[25] Jackson, J.D. 1975, {\it Classical Electrodynamics}, (Wiley, New
York), eqs (4.42) and (5.88).

[26] Dray, T. and Joshi, P.S. 1990, {\it Class. Q. Grav.} {\bf 7},
 41-9.

[27] Sumeruk, A. 1993, {Master's Thesis} University of Cape Town.

[28] Sumeruk, A. Hellaby, C. and Ellis, G.F.R. 1993, {\it in
preparation}.

[29] Smolin, L. 1992, {\it Class. Q. Grav.} {\bf 9}, 173.

 }

 \end{document}